\documentclass[prb,preprint]{revtex4-1} 

\usepackage{amsmath}  
\usepackage{amsfonts} 
\usepackage{graphicx} 

\begin{document}


\title{Incorporating the Stern-Gerlach delayed-choice quantum eraser into the undergraduate quantum mechanics curriculum}

\author{William F. Courtney}
\affiliation{2801 Montair Ave., Long Beach, CA 90815-1054 USA}
\author{Lucas B. Vieira}
\affiliation{Universidade Federal de Minas Gerais, 6627 Av. Ant\^onio Carlos, Belo Horizonte, MG, Brazil 31270-901}

\author{Paul S. Julienne}
\affiliation{Joint Quantum Institute, National Institute of Standards and Technology, and University of Maryland, 100 Bureau Drive, Stop 8423,
Gaithersburg, MD 20899-8423, USA}

\author{James K. Freericks}
\affiliation{Department of Physics, Georgetown University, 37th and O Sts. NW, Washington, DC 20057 USA}


\date{\today}

\begin{abstract}
As ``Stern-Gerlach first'' becomes the new paradigm within the undergraduate quantum mechanics curriculum, we show how one can extend the treatment found in conventional textbooks to cover some of the exciting new developments within the quantum field. Namely, we illustrate how one can employ Dirac notation and conventional quantum rules to describe a delayed choice variant of the quantum eraser which is realized within the Stern-Gerlach framework. Covering this material, allows the instructor to reinforce notions of changes of basis functions, quantum superpositions, quantum measurement, and the complementarity principle as expressed in whether we know ``which-way'' information or not. It also allows the instructor to dispel common misconceptions of when a measurement occurs and when a system is in a superposition of states. We comment at the end how a similar methodology can be employed when the more conventional two-slit experiment is treated.
\end{abstract}

\maketitle 

\section{Introduction} 

The Stern-Gerlach experiment was originally performed\cite{stern_gerlach} by Otto Stern and Walter Gerlach in 1922. While it can be thought of as simply an experiment to separate an atomic beam into its different projections of angular momentum, it also illustrates a number of different quantum phenomena. One can use it to show that quantum phenomena require a probabalistic interpretation. One can use it to show that quantum states cannot have definite projections of angular momentum on two non-collinear axes. It also acts as one of the simplest paradigms of a two-state quantum system (for the case of a spin-one-half atom like silver), illustrating the discreteness of quantum eigenvalues.

Educators have long realized the importance of this experiment. It has appeared in many textbooks. Here, we highlight a few texts that bring this experiment to the forefront, by employing it as the first, or as one of the first, quantum experiments that the student encounters. These texts deviate from the far more common norm of covering quantum mechanics from a historical perspective. We believe that there are significant advantages to proceeding in this ``Stern-Gerlach first'' methodology, as it allows the students to encounter experiments that they can easily analyze right from the beginning. Furthermore, as we show here, one can extend those treatments to allow the students to encounter sophisticated quantum paradoxes even before they learn what a coordinate-space wavefunction is.

{\it The Feynman Lectures on Physics},\cite{feynman} introduces the Stern-Gerlach experiment quite early in its discussion of quantum mechanics, actually covering the spin-one case before the spin-one-half case. This text also describes what we will call the Stern-Gerlach analyzer loop (following Styer, see below); this device is sometimes called a Stern-Gerlach quantum eraser by other authors, but we will be reserving that language for the more complex eraser we describe below. Sakurai employed the Stern-Gerlach experiment early in his textbook\cite{sakurai} and used it to also discuss the Bell experiments. Our treatment of the subject is influenced most by Styer's wonderful text {\it The Strange World of Quantum Mechanics},\cite{styer} which introduces a number of complex ideas including the two-slit experiment, Wheeler's delayed choice, the Einstein-Podolsky-Rosen paradox, and the Bell experiments all within the framework of the Stern-Gerlach experiment. His text also carefully describes the classical version of the experiment, which is critical for students to master in order to appreciate the quantum nature of the real experiment. 

Two recent undergraduate textbooks, Townsend's {\it A Modern Approach to Quantum Mechanics}\cite{townshend} and McIntyre, Manogue and Tate's {\it Quantum Mechanics: A Paradigms Approach}\cite{mcintyre} both adopt the Stern-Gerlach first paradigm, introducing students to this experiment as their initial encounter with quantum mechanics. While both texts move on to a more conventional style of quantum treatment afterwards, this critical change allows students to dive into a quantum system that they can understand all aspects of and allows them to lean on this knowledge as they learn about new and different quantum phenomena in the remainder of the books.

The quantum field has also seen numerous developments that have not yet made it into most introductory quantum texts. For example, in the 1980's, John Wheeler introduced the notion of delayed choice,\cite{wheeler1,wheeler2} where an experimental apparatus is modified {\it while the particle is moving through it}, in such a way that the modification determines what type of measurement will be performed. Wheeler hypothesized that these types of experiments, which can differentiate whether a particle goes through just one slit, or two slits at the same time, in a two-slit experiment, have the spooky behavior of acting like the quantum particle is able to influence what has already occurred, by going backwards in time. It turns out that this awkward notion is easily dispelled when one properly interprets when the system is in a superposition of states and when a measurement collapses the wavefunction.\cite{philosophy}~~Nevertheless, the notion of a delayed choice experiment being employed to change the outcome is a remarkably powerful demonstration, as can be seen by numerous videos available on the internet which illustrate this phenomena using crossed polarizers over each slit of the two-slit experiment and an additional polarizer, whose orientation can be rotated, just before the light hits the detector screen. Those videos are actually showing a delayed-choice quantum-eraser variant, which we describe next.

The quantum eraser idea of Scully, Englert, and Walther,\cite{scully} is even more bizarre. Here, what is generally done is that the particles that are input into a two-slit experiment  (or a Mach-Zehnder interferometer) are also entangled with other quantum particles, which can be employed to provide which-way information. As long as the entanglement persists, the conventional interference effects are suppressed. But if the entanglement is removed, then the interference effects also return. What is remarkable about these experiments is that they often can have the choice for whether we see the interference or not be decided well after the quantum particles have gone through the device. In some sense, one can think of the delayed-choice aspect as providing a filter which removes the results of the experiment that do not provide the interference one is trying to restore. In this sense, which we develop more fully below, the interference is never fully restored, because the entanglement and subsequent filtering always remove some particles from the experiment, so the interference oscillations have a smaller amplitude than what one would see if there was never any entanglement in the first place. Indeed, it is an interesting open question whether interaction-free measurements could be coupled with delayed-choice quantum erasers to restore all or nearly all of the full amplitude of the quantum interference after the eraser is employed.

In the future, we believe that more and more students will be exposed to the  Stern-Gerlach experiment early in the quantum curriculum. It is for this reason that we want to show how one can employ these experiments to cover quite advanced, and frankly quite bizarre phenomena, early on in a course. Mastering this material will instill in the students the confidence needed to be successful in the remainder of their quantum class, but it also will allow them to experience some of the truly strange behavior that lies within quantum mechanics, and to know that it can be quantitatively described within the theory.

\section{Preliminaries for the Stern-Gerlach experiment}

Three of us have been involved in a MOOC entitled {\it Quantum Mechanics For Everyone} which is running on Edx from April 2017 until March 2019.\cite{qmfe}~~Freericks is the lead instructor and course developer, Vieira created over half of the computer-based tutorials that run under JavaScript,\cite{javascript} and Courtney was in the original student cohort. We describe how the MOOC covers the Stern-Gerlach experiment to define the terminology and to introduce the different devices we need to describe the delayed-choice quantum-eraser variant. As mentioned above, this treatment is heavily influenced by both Styer's\cite{styer} approach and Feynman's.\cite{feynman}

To begin, students need to understand how a classical Stern-Gerlach experiment would work. As Styer shows, one can develop that a current loop precesses in a field, with a constant projection on the field axis, and it feels a force if the field is inhomogeneous in space. It is important that the students recognize that one needs an inhomogeneous field to apply a force proportional to the projection, and that the projection does not change during the time the current loop is in the field.

\begin{figure}[htb]
\centering
\includegraphics[width=5.0in]{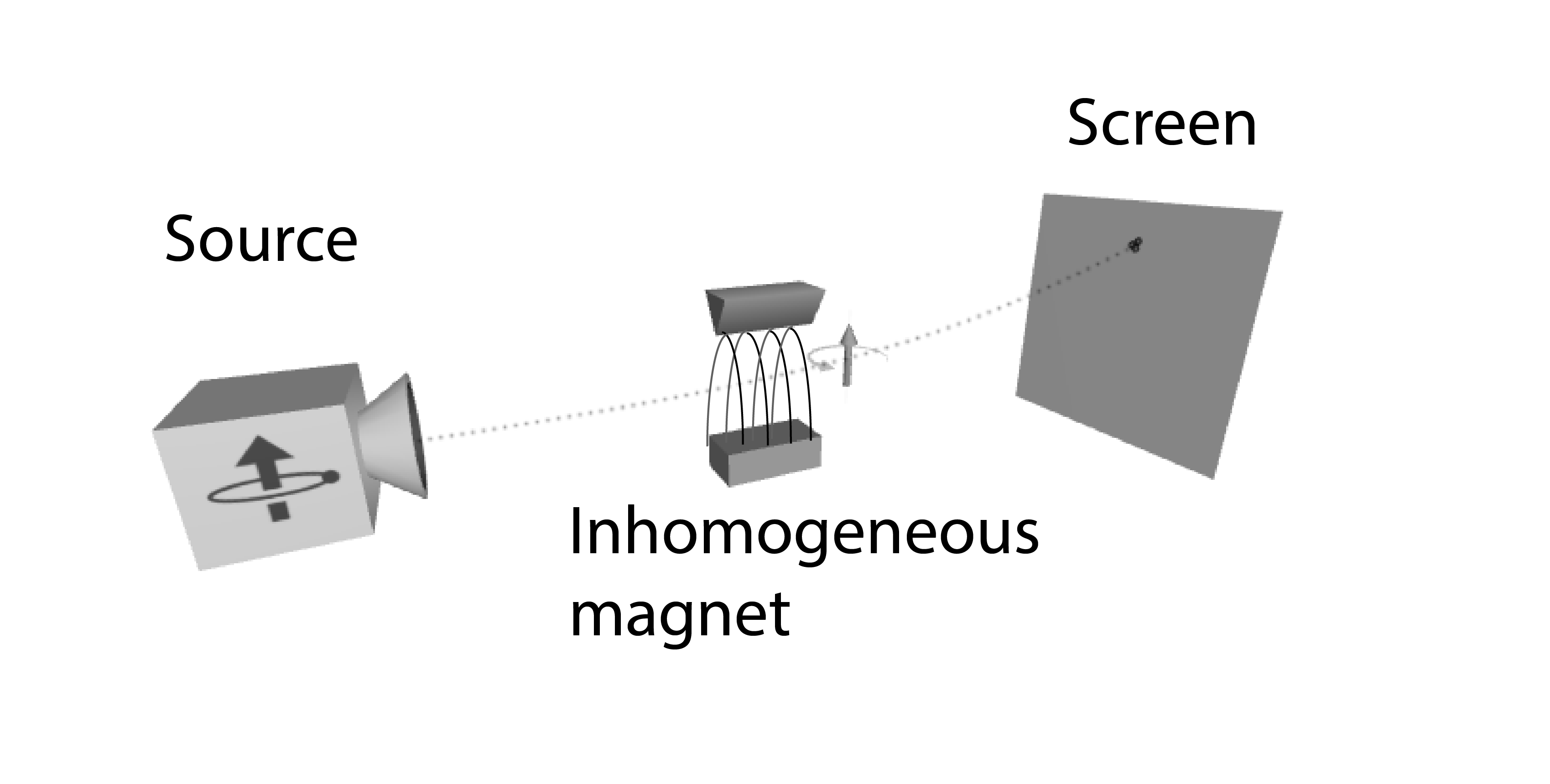}
\caption{Schematic of the classical Stern-Gerlach experiment, with a source of classical current loops, an inhomogeneous magnetic field occuring between the shaped magnetic poles with field lines sketched, and a screen to detect the projection of the current loops as they move through the device. The curved dashed line indicates the current loop trajectory. This current loop has a maximal projection on the z-axis, so it is deflected upwards.}
\label{fig: classical_sg}
\end{figure}

By using a beam of atoms shot through an inhomogeneous field, one now can describe how the device will separate the beam into different projections of the angular momentum onto the field axis, with the spatial position correlated with the magnitude of the projection of the angular momentum. One can describe such an experiment as analogous to a triangular prism, which separates white light according to its color. An example of such a classical Stern-Gerlach experiment, similar to what is used in the MOOC, is shown in Fig.~\ref{fig: classical_sg}.

\begin{figure}[htb]
\centering
\includegraphics[width=2.75in]{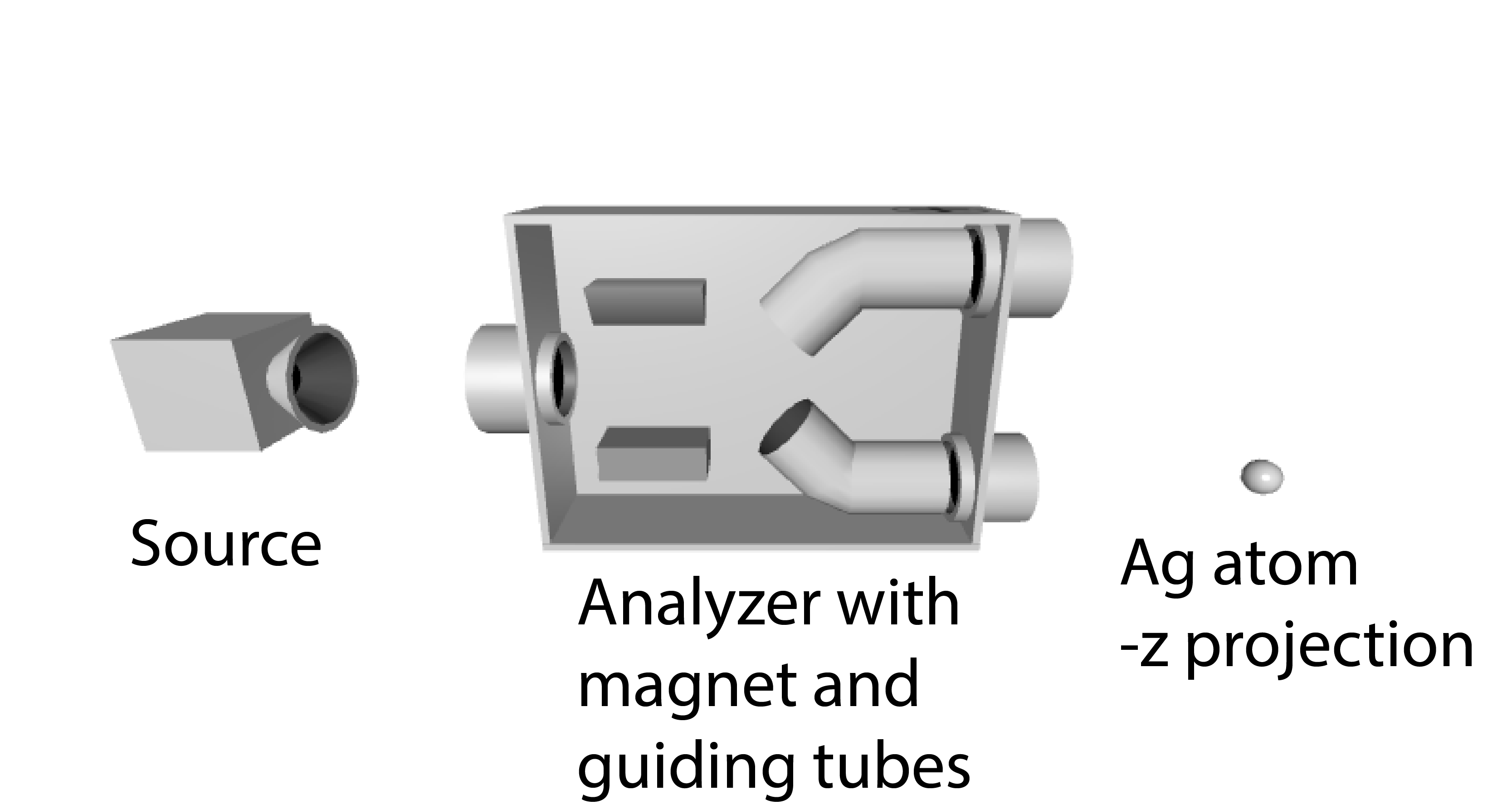}
\includegraphics[width=2.75in]{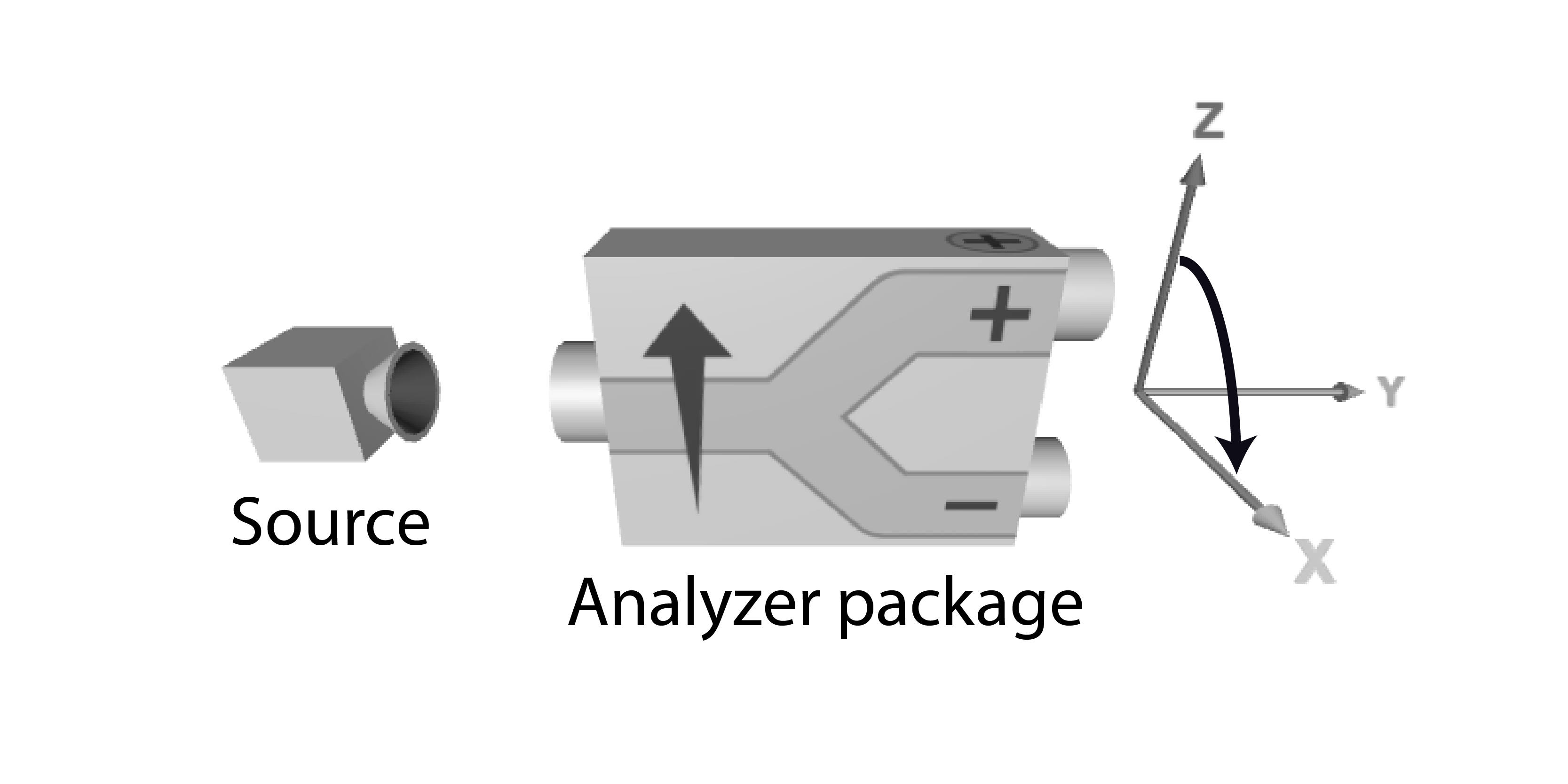}
\caption{Schematic of the quantum Stern-Gerlach experiment with silver atoms, which produces  only two deflections. The packaging with an inhomogeneous magnet and ``guiding tubes'' on the left is covered with a schematic annotation creating the Stern-Gerlach analyzer on the right, which also illustrates the coordinate system used to describe the orientation of the analyzer in the $x-z$ plane. }
\label{fig: quantum_sg}
\end{figure}

Of course, the quantum experiment does not provide a continuous beam of separated projections. When run using silver atoms, it shows just two different projections of the angular momentum: one corresponding to $+\frac{1}{2}\mu_B$ and one to $-\frac{1}{2}\mu_B$, with $\mu_B$ the Bohr magneton. This quantum result motivates a number of follow-up experiments to understand this phenomena. But before describing them briefly, we need to show how one packages the Stern-Gerlach analyzer for use in further thought experiments (see Fig.~\ref{fig: quantum_sg}). Since the quantum Stern-Gerlach experiment on silver produces only two results, regardless of the orientation of the analyzer, we think of the experiment as a separation region where the magnets are positioned and ``tubes'' that collect the atoms according to their projections and direct them to the respective + and -- exits (curving their velocities to be horizontal). The device is packaged together so that we have a direction of the field given by the arrow, the sense of the inhomogeneity of the field also given by the widening of the arrow's shaft, and the two exits.
The tubes that curve to the exits can be thought of as being constructed from an inhomogeneous magnet oriented opposite to the initial separating magnet, which curves the paths to be horizontal and ejects the atomic beams in a horizontal direction after emerging from the analyzer.

\begin{figure}[htb]
\centering
\includegraphics[width=1.85in]{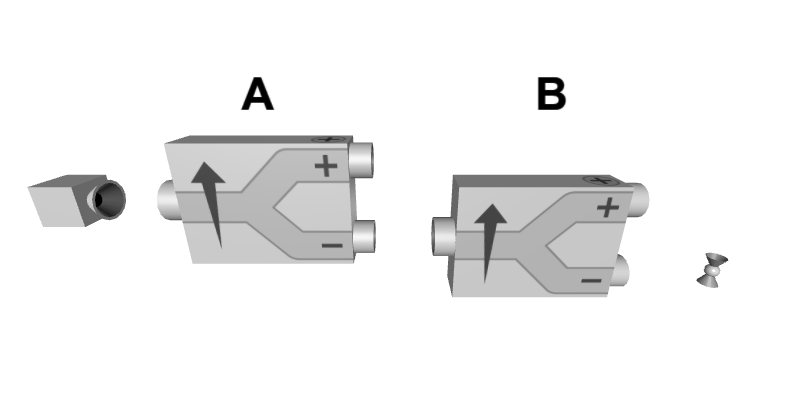}
\includegraphics[width=1.85in]{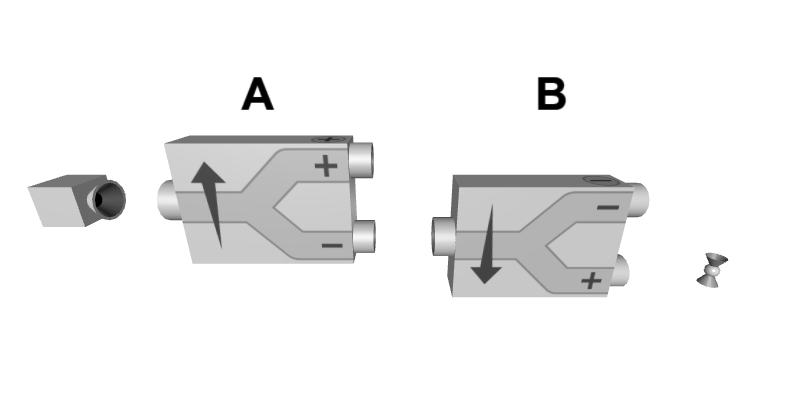}
\includegraphics[width=2.0in]{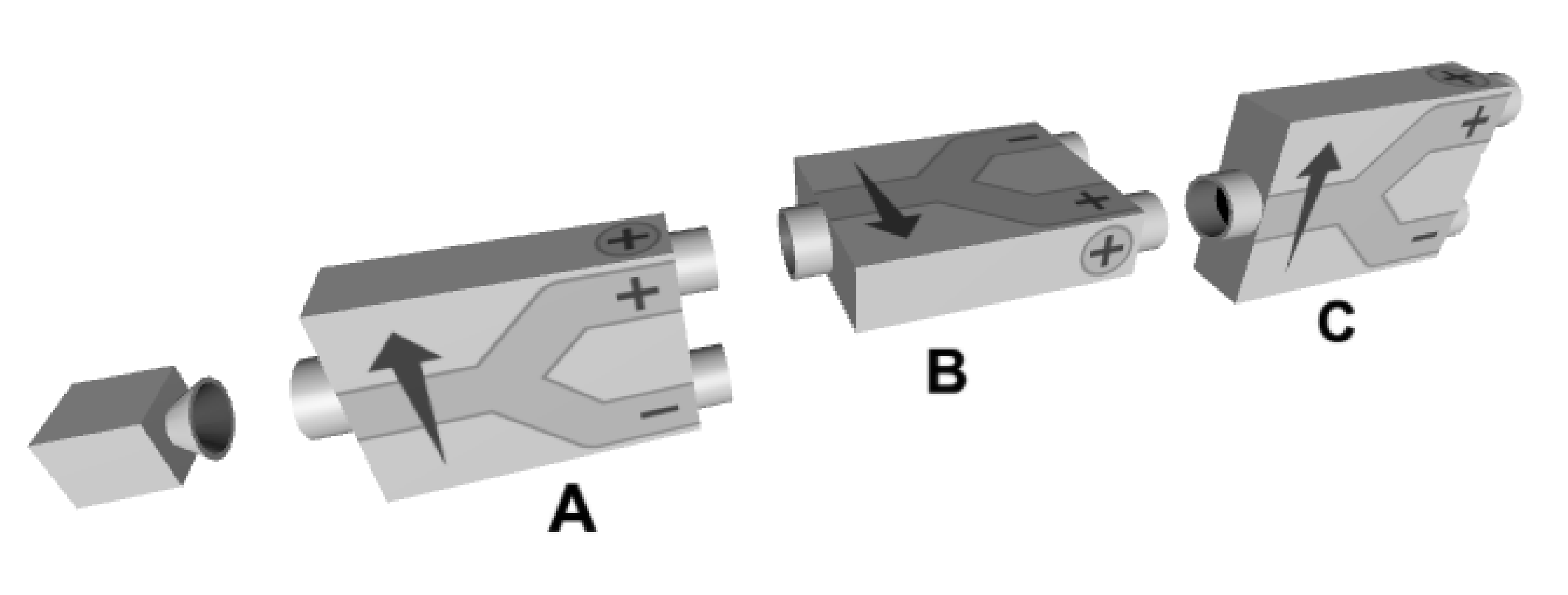}
\caption{Three different experiments with Stern-Gerlach analyzers.  (Left), measure on the $z$-axis (A), then measure on the $z$-axis again (B). This experiment shows that the analyzers are reproducible. (Center), measure on the $z$-axis (A), then on the $-z$-axis (B). This experiment shows that a positive projection on one axis is a negative projection on the inverse axis and {\it vice versa}. (Right), measure on $z$ (A), measure on $x$ (B), then measure on $z$ again (C). Since we only input atoms with a positive $z$-axis projection into the horizontal analyzer, one might expect  that they will all emerge with a positive projection through the third analyzer, but we find half of the time they are positive and half of the time negative, because the angular momentum operators in different Cartesian directions are incompatible operators. }
\label{fig: expts}
\end{figure}

The Stern-Gerlach analyer can then be employed in a series of experiments (see Fig.~\ref{fig: expts}). If we measure on $z$ and on $z$ again (left panel), we see the results are reproducible. If we measure on $z$ and then on $-z$, we see the relationship between measuring on axes oriented oppositely to each other (center panel). If we measure on $z$, then on $x$, and then on $z$ again, we see that atoms can only have a projection on the last axis on which they were measured (right panel). In other words, if the atoms always enter the horizontal analyzer with a positive projection on the $z$-axis, they can emerge from the final analyzer either from the + or -- exit of the vertical analyzer.  These experiments show how one must invoke a probabilistic interpretation, because we cannot foretell the outcome of any single experiment, only the probability after many have been performed. They also show that incompatible operators cannot have simultaneous eigenvalues, as we cannot have a state with a definite $z$-axis projection {\it and} a definite $x$-axis projection.

The formalism to describe these three experiments is straightforward. We employ Dirac bra-ket notation, where a bra $\langle \psi |$ and a ket $|\psi\rangle$ are the notations for a quantum state $\psi$. Forming a bracket, such as $\langle \psi' | \psi \rangle$ corresponds to the inner product between the two different states. One can simply think of the bra and the ket as being place holders for the labels that denote the different states. 

In order to describe the experiments, we need just three postulates: (i) the norm of all quantum states is 1, so $\langle\psi |\psi\rangle=1$; (ii) the measurement by a Stern-Gerlach analyzer corresponds to a projection onto the state corresponding to the exit of the analyzer (for example, $|\uparrow\rangle_z \,_z\langle \uparrow|$ is the projector onto the positive projection atomic state along the $z$-axis); and (iii) the modulus squared of the final projected wavefunction yields the probability. Note that all quantum states are unit norm, but a projected wavefunction corresponds to a quantum state multiplied by a scalar whose magnitude is less than or equal to one.

Using this formalism, we have for experiment 1 (Fig.~\ref{fig: expts}, left) the following analysis. The initial state entering the second analyzer is $|\downarrow\rangle_z$. After passing through that  analyzer, it is projected to the wavefunction $|\downarrow\rangle_z \,_z\langle \downarrow|\downarrow\rangle_z=|\downarrow\rangle_z$, because $_z\langle \downarrow|\downarrow\rangle_z=1$. Squaring, gives a probability of $\,_z\langle\downarrow|\downarrow\rangle_z=1$.

Using the identities that $|\uparrow\rangle_z=|\downarrow\rangle_{-z}$ and $|\downarrow\rangle_z=|\uparrow\rangle_{-z}$, allows us to analyze experiment 2 (Fig.~\ref{fig: expts}, center). The wavefunction after emerging through the first analyzer is $|\downarrow\rangle_z$, because we examine only the atoms exiting the $-$ exit. Then
we find we need to evaluate $|\uparrow\rangle_{-z} \,_{-z}\langle \uparrow|\downarrow\rangle_z$. Replacing the states labeled on the $-z$ axis, by the $z$-axis counterparts, yields $|\downarrow\rangle_z \,_z\langle\downarrow|\downarrow\rangle_z=|\downarrow\rangle_z$. Squaring gives a probability of 1, hence all atoms exit the + exit of the second analyzer.

For the last experiment (Fig.~\ref{fig: expts}, right), we need to know the representation of the $x$-states in terms of the $z$-states: $|\uparrow\rangle_x=\frac{1}{\sqrt{2}}(|\uparrow\rangle_z+|\downarrow\rangle_z)$, which either can be simply told to the students, or which can be easily developed if one introduces the spin operators and their properties. Then we have that the wavefunction of the system after exiting the first analyzer is $|\uparrow\rangle_z$. The wavefunction exiting the + exit of the $x$-axis analyzer is then $|\uparrow\rangle_x\,_x\langle\uparrow|\uparrow\rangle_z=\frac{1}{2}(|\uparrow\rangle_z+|\downarrow\rangle_z)(_z\langle\uparrow|+_z\langle\downarrow|)|\uparrow\rangle_z$. Using the fact that $\,_z\langle\uparrow|\downarrow\rangle_z=0$, then yields the output wavefunction as $\frac{1}{2}(|\uparrow\rangle_z+|\downarrow\rangle_z)$. After being measured in the final analyzer, we project to the wavefunction $\frac{1}{2}|\uparrow\rangle_z \,_z\langle\uparrow|(|\uparrow\rangle_z+|\downarrow\rangle_z)=\frac{1}{2}|\uparrow\rangle_z$. So the probability to emerge from the + exit of the third analyzer is $\frac{1}{4}\,_z\langle\uparrow|\uparrow\rangle_z=\frac{1}{4}$. The same probability occurs for exiting the -- exit of the last analyzer. One way of summarizing this behavior is to say that {\it the atom is stupid}---implying it only remembers the last axis it was projected onto. Hence, an atom entering with a positive vertical projection, will then assume a horizontal projection, if measured on the $x$-axis, and thereafter may have a negative projection on the vertical axis if measured on the $z$-axis. This is because the atom cannot have a definite projection on the $x$- and $z$-axes at the same time. What about the total probability? If only 25\% exit the + exit and 25\% exit the -- exit, we have lost 50\% of the atoms. Indeed, we have, as those atoms emerged from the $-x$ exit of the horizontal analyzer and were ignored in the experiment.

Next, we describe the Stern-Gerlach analyzer loop (also sometimes called the Stern-Gerlach eraser, but we use that term for a different variant).  This device nominally splits the atomic beam according to its projection along the orientation of the analyzer loop and then rejoins it again. But there is no way for us to verify this behavior unless a measurement is performed, so it is safer to  say that the analyzer loop allows us to measure the projection of an atom in the analyzer loop orientation if we choose to, or to leave the atom in its original state if we choose not to perform a measurement. (This issue is similar to the situation in a two-slit experiment where we do not know which slit the photon goes through or how it ``interferes with itself'' if we do not watch at the slits.) Because we created the Stern-Gerlach analyzer to pipe the atoms into horizontal beams at the exit, we merely need to attach two oppositely oriented analyzers back-to-back in order to make the analyzer loop (see Fig.~\ref{fig: analyzer_loop}). As we will see below, we also could call this a ``measurable basis-changer,'' but we stick with the original name from Styer.

\begin{figure}[htb]
\centering
\includegraphics[width=3.10in]{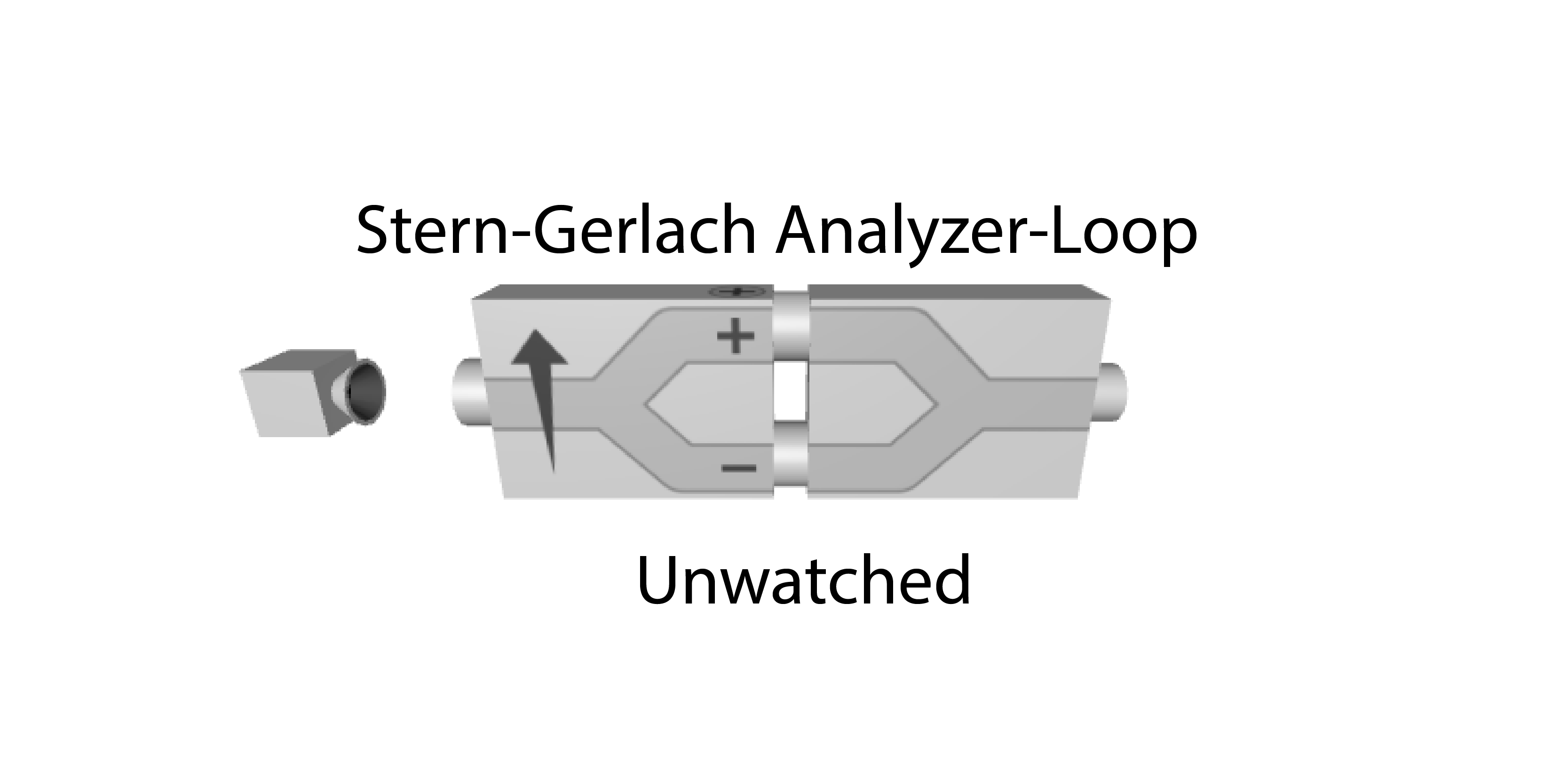}
\includegraphics[width=3.25in]{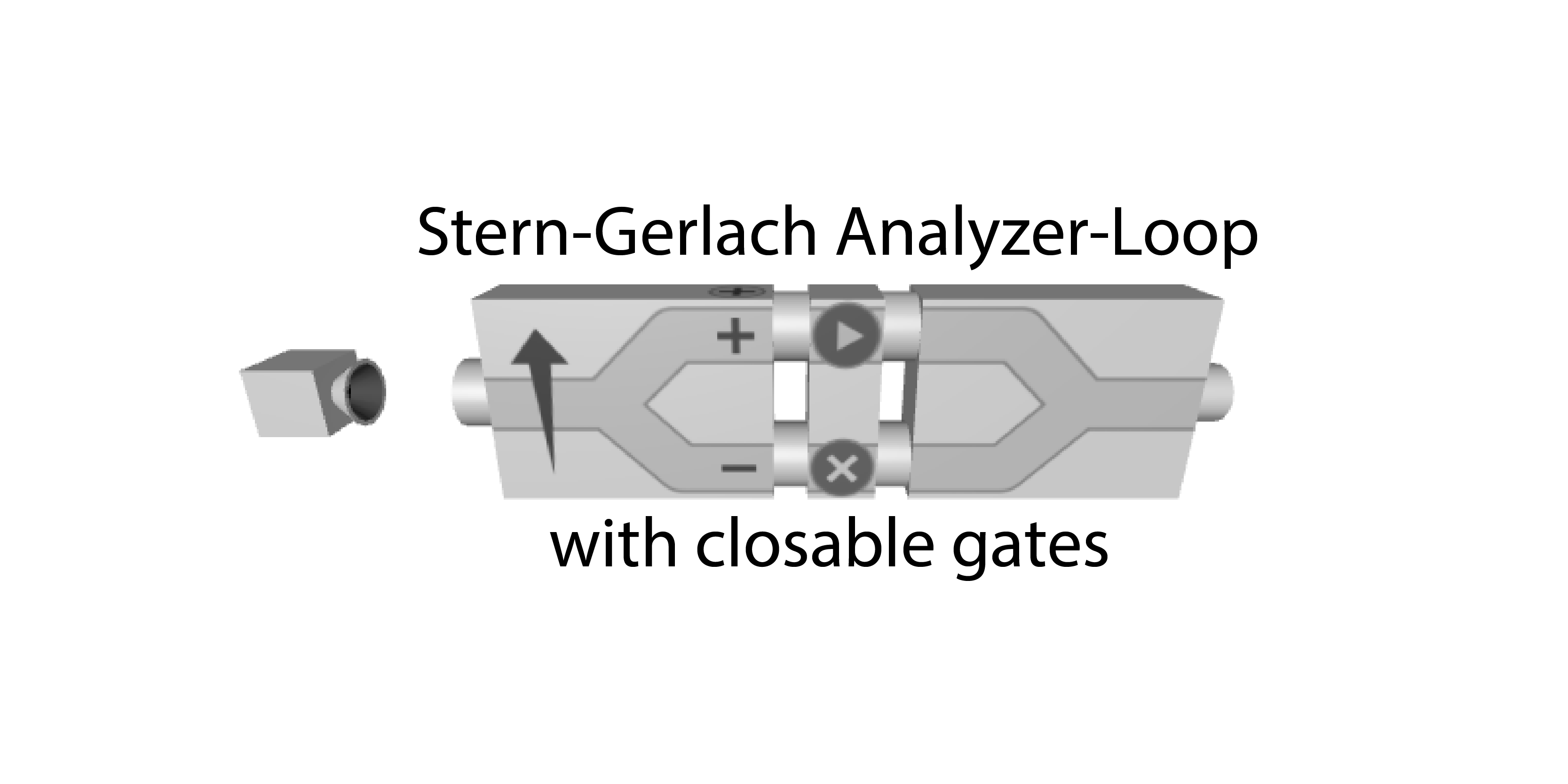}
\caption{(Left) Schematic of an analyzer loop, which can be thought of as two Stern-Gerlach analyzers attached back-to-back. If no measurement is made, then the analyzer loop does not alter the quantum state of the atom and it emerges with the same state it entered. If one of the paths is blocked, the atom emerges with the state given by the path that is not blocked. (Right) a Stern-Gerlach analyzer loop with a flow-through gate. The gate can be independently controlled to block zero, one, or two branches of the analyzer loop. The pictured flow-through gate is configured to block the lower branch of the analyzer loop (as indicated by the $\times$).}
\label{fig: analyzer_loop}
\end{figure}

Instead of thinking of the analyzer loop as separating and rejoining the atomic beams, since this {\it is not a measurement}, it is better to view the analyzer loop as placing the atoms into a superposition of states according to the orientation of the analyzer loop. If no measurement is made, the original state of the atom entered with is unchanged. It is easiest to describe this as the situation where {\it the analyzer loop does nothing}.  If, on the other hand we block one of the analyzer loop paths, then the atom is projected onto the state that {\it was not} blocked. The notion of an eraser that other authors use follows from the fact that the analyzer loop creates a superposition of states, but then restores the original state if no path is blocked. But we aver that a better way to describe this situation is that the analyzer loop acts as a {\it change of basis} from whatever initial basis state the atom enters the analyzer loop into the basis corresponding to the axis oriented in the direction of the analyzer loop and then back to the original basis if no measurement is made. For example, if the atom starts in a down state along the $x$-axis, enters an analyzer loop oriented along the $z$-axis, then the atomic state can be thought of as initially being in the state $|\downarrow\rangle_x$, then being expressed in the $z$-basis as $\frac{1}{\sqrt{2}}(|\uparrow\rangle_z-|\downarrow\rangle_z)$ when the atomic beam ``splits into two branches,'' and finally, emerging as $|\downarrow\rangle_x$ after the ``beams rejoined.'' Of course, this means nothing happened to the atom, because the quantum state remained the same regardless of what basis it was expressed in. It is important to realize that the state does not collapse unless a measurement is made {\it inside} the analyzer loop (by blocking one path, for example). We feel this point is an important one to make with students, because the notion of a state and the notion of the basis chosen to represent the state are often confused by students.  The analyzer loop provides a unique opportunity to properly describe this subtle distinction.

\begin{figure}[htb]
\centering
\includegraphics[width=5.0in]{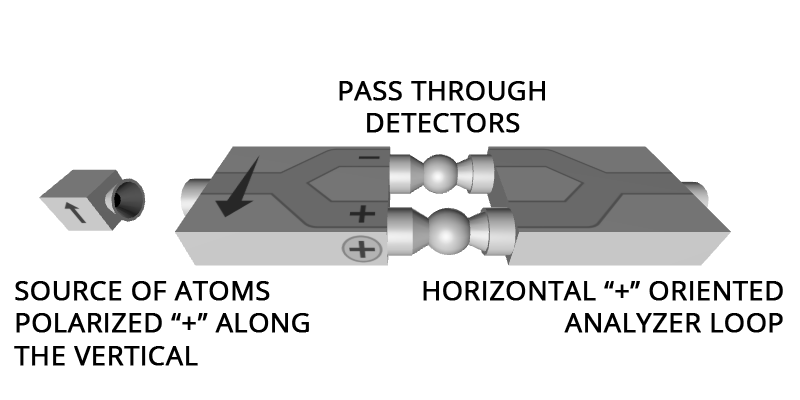}
\caption{Analyzer loop with pass-through detectors, which allow the path to be watched as the atoms move on the + or -- branches. We always see a full atom on one branch or the other. Watching the atoms changes their output state because it acts just like a measurement. }
\label{fig: analyzer_loop_pass_through_detector}
\end{figure}

If, however, we watch at the branches with a device called a pass-through detector, shown in Fig.~\ref{fig: analyzer_loop_pass_through_detector}, then we are performing a measurement, and the results of the experiment will change. For example, consider the arrangement given in Fig.~\ref{fig: analyzer_loop_pass_through_detector}. The analyzer loop has a $|\uparrow\rangle_z$ state input. When an atom passes through one of the arms of the horizontal analyzer loop, it is measured by the pass-through detector. This corresponds to a projection onto the $x$-axis via $|\uparrow\rangle_x \,_x\langle \uparrow |$  when detected on the + branch or via $|\downarrow\rangle_x \,_x\langle \downarrow |$ when detected on the -- branch. If we see an atom on the + branch, then we find the measurement due to the pass-through detector implies we have the wavefunction $|\uparrow\rangle_x \,_x\langle\uparrow|\uparrow\rangle_z=\frac{1}{\sqrt{2}}|\uparrow\rangle_x$ emerge from the exit of the analyzer loop. Similarly, if the atom passes through the -- branch, we have the down wavefunction along the $x$-axis. Half of the time, we obtain an up spin and half of the time a down spin along the vertical axis. So watching at the two branches is the same as measuring along them, because it provides us with which-way information. Note that at no time do we see half of an atom going on two different paths. We always see a full atom on one path or on another path.

\section{Delayed-Choice Quantum-Eraser Stern-Gerlach Experiment}

\begin{figure}[htb]
\centering
\includegraphics[width=3.0in]{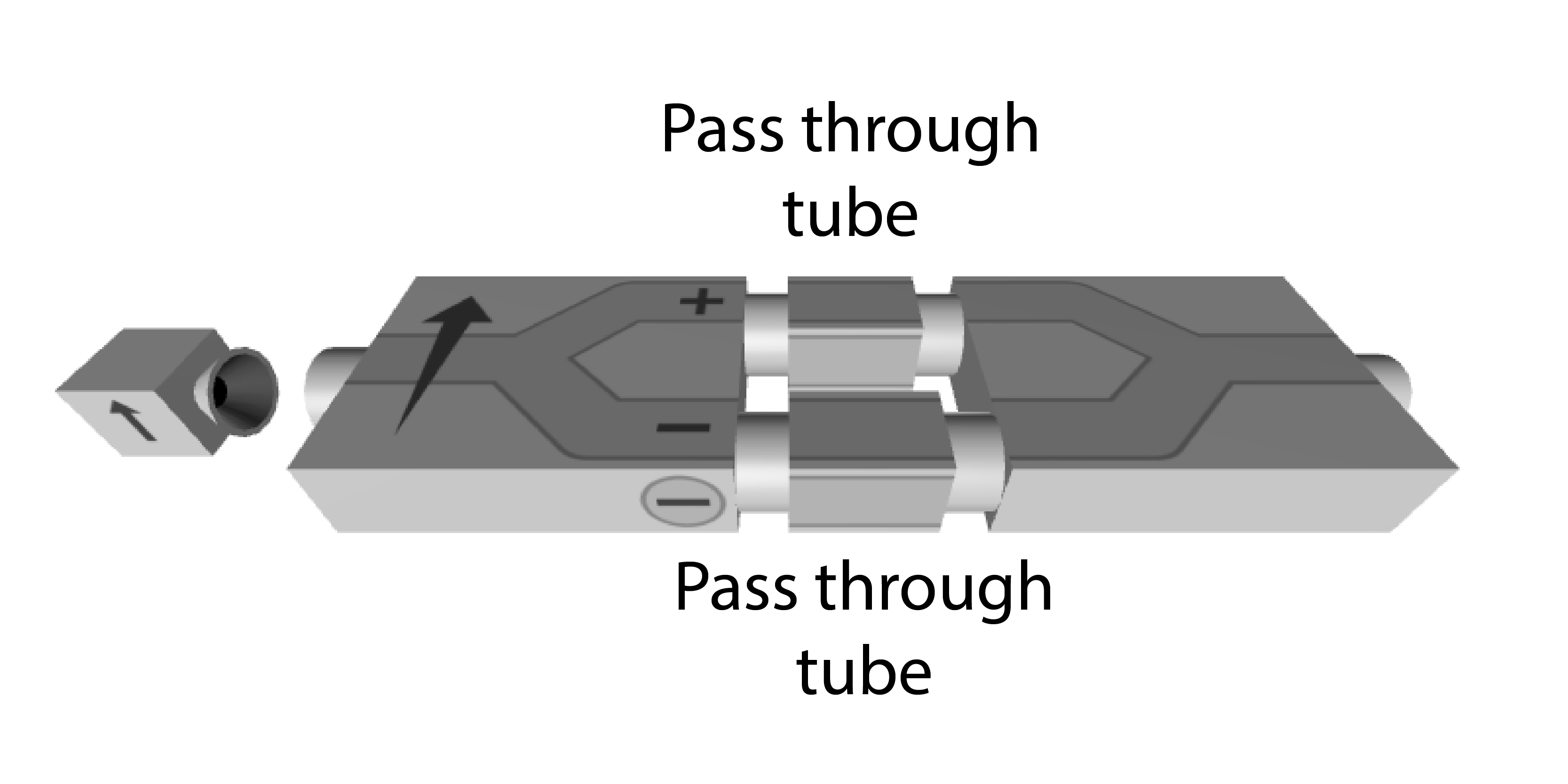}
\includegraphics[width=3.0in]{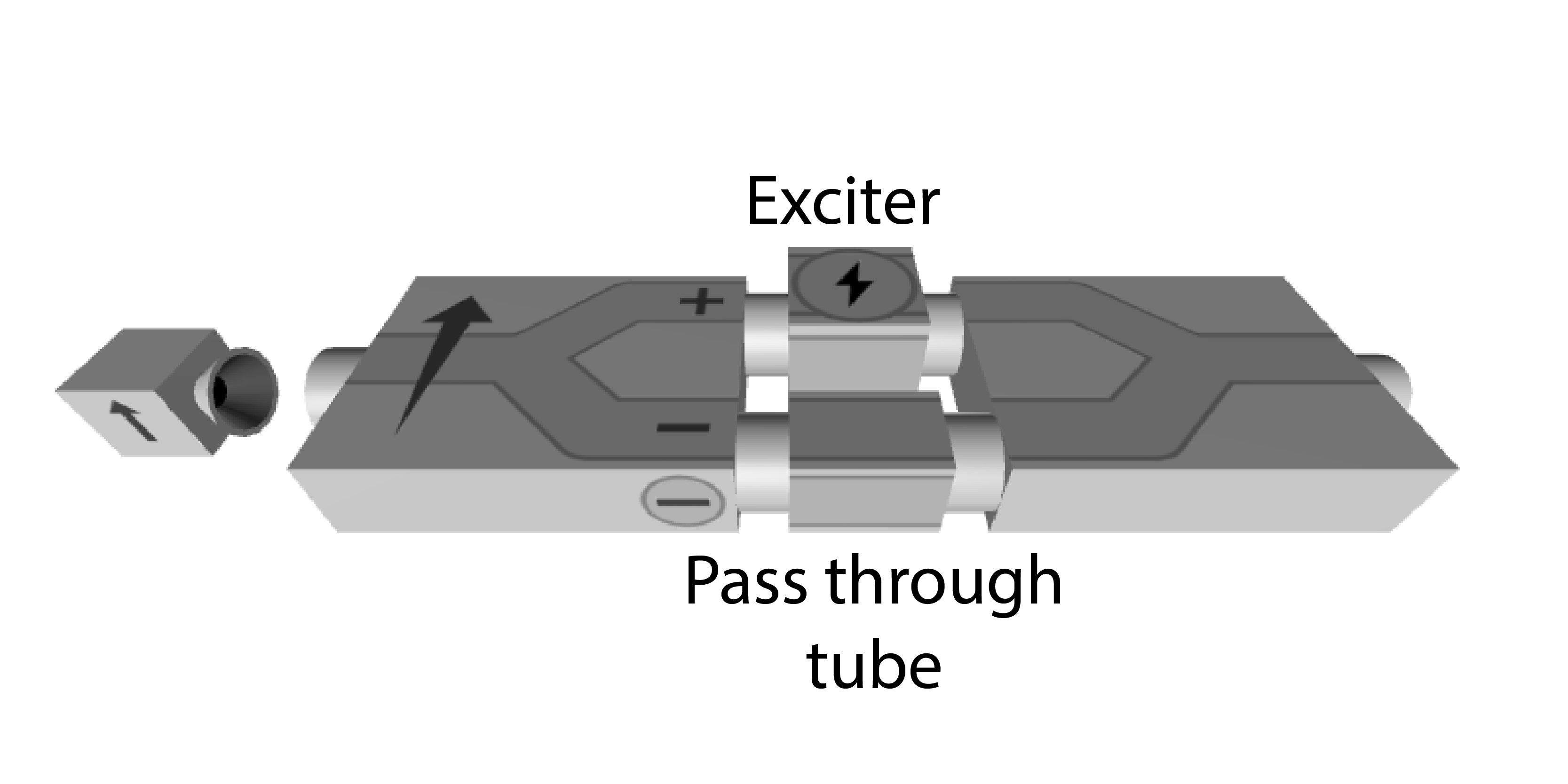}
\caption{(Left) Analyzer loop with pass-through tubes, which allow the beam to pass without blocking a path or detecting if an atom went through a path. (Right) Analyzer loop with an exciter on the + branch and a pass through tube on the -- branch. }
\label{fig: analyzer_loop_exciter} 
\end{figure}

We begin by re-iterating the quantum superposition effect of the analyzer loop. We start with an input atom in a definite state. The analyzer loop re-expresses the atom in a superposition of states according to the basis directed along the orientation of the analyzer loop. It then re-expresses the atomic state in the original basis as it emerges from the analyzer loop. This analog of quantum interference effects corresponds to the fact that the atoms all emerge in the same state they entered even though they were expressed as a superposition along a different axis when they were inside the apparatus. Since a basis change does not change the underlying quantum state, the unmeasured analyzer loop effectively does nothing to the atom.

We are now ready to start discussing the quantum eraser. The eraser works by first tagging the atoms via their internal quantum numbers, which acts in many respects like a measurement when the atoms are on one of the two analyzer loop branches. But, the tagging procedure still leaves the atoms in a pure superposition of quantum states, so a measurement via a projection has not yet been made. For example, we assume there are two internal states, {\it unrelated to the spin of the atom}, which can be excited or de-excited. We attach an exciter to the + branch, as depicted in Fig.~\ref{fig: analyzer_loop_exciter}, right and denoted with the lightning bolt symbol. This device excites the internal structure of the atom from the ground state to the excited state without affecting the spin structure. This then can be employed to determine which path of the analyzer loop the atom takes simply by measuring the internal state of the atom.

Hence, tagging the atoms on the + branch by exciting them allows us to determine ``which-way'' information. We have correlated the internal state of the atom with the spin projection along the $x$-axis. If we measure the internal state, then we know which path it took through the analyzer loop, in direct analogy to what happened when we watched at the arms of the analyzer loop with pass-through detectors.  But it is not exactly the same, because {\it we have not yet performed the measurement}. Our system has only been transformed to a superposition of states at this stage. We must use a direct product notation to describe this. We let $|ES\rangle$ denote the excited internal state and $|GS\rangle$ denote the internal ground state. Then the exciter will take an input state of $|\uparrow\rangle_z\otimes |GS\rangle=\frac{1}{\sqrt{2}}(|\uparrow\rangle_x+|\downarrow\rangle_x)\otimes|GS\rangle$ and transform it to $\frac{1}{\sqrt{2}}(|\uparrow\rangle_x\otimes|ES\rangle +|\downarrow\rangle_x\otimes |GS\rangle)$, which is a superposition corresponding to a pure (but entangled) quantum state. But, by measuring the internal state of the atom, we can determine which branch the atom took through the analyzer loop and hence we know the projection of its spin, even if we do not directly measure the projection of the spin. In any case, we cannot immediately restore the initial spin state of $|\uparrow\rangle_z$ because of the complex nature of the superposition, which has entangled the internal degrees of freedom of the atom with the different spin states. We successfully erase this information when we restore the atoms to their original state (ground state and original spin projection). Hence tagging is not the same as a measurement, because the system is only placed into a superposition of states and the measurement has not yet been performed. This gives us the possibility to ``untag'' the atoms and restore the original state because no measurement has occurred. This untagging procedure is called a quantum eraser.

\begin{figure}[htb]
\centering
\includegraphics[width=5.0in]{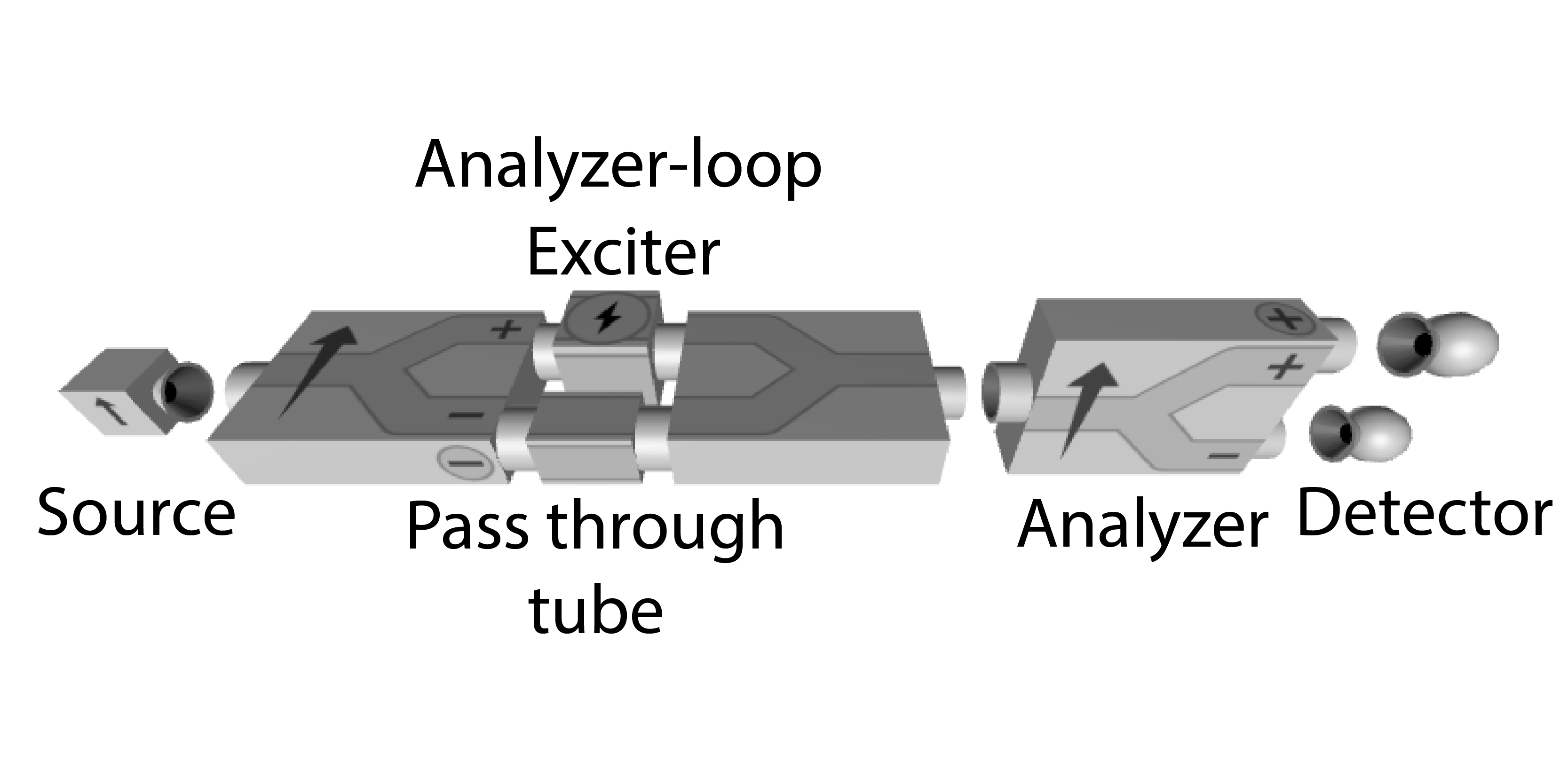}
\caption{ First stage of the eraser experiment with an analyzer loop that has an exciter on the + path and a vertical analyzer to detect the atoms at the end of the experiment. }
\label{fig: analyzer_loop_eraser}
\end{figure}

If we extend the analyzer-loop with exciter experiment by having the analyzer-loop output go through a vertical analyzer loop, as shown in Fig.~\ref{fig: analyzer_loop_eraser}, we will find that the exiting atoms will emerge half of the time from the + branch and half of the time from the -- branch. In addition, the atom will be in the ground state half of the time and in the excited state half of the time, {\it  with no correlation between the spin state and the internal state after emerging from the vertical Stern-Gerlach analyzer.}
Nevertheless, by measuring the internal state of the atom, we can immediately know whether it went through the + or -- branch of the analyzer loop, even though we have scrambled the spin projection, by measuring it on the $z$-axis.

Let's be sure we understand this by carefully going through the analysis. We measure the probability to exit the + exit of the vertical analyzer by projecting the output of the analyzer loop onto $|\uparrow\rangle_z \,_z\langle \uparrow|$ and then finding the norm of the final projected wavefunction. Hence, the projection produces
\begin{equation}
\frac{1}{\sqrt{2}}(|\uparrow\rangle_z \,_z\langle\uparrow|\uparrow\rangle_x\otimes|ES\rangle+|\uparrow\rangle_z \,_z\langle\uparrow|\downarrow\rangle_x\otimes|GS\rangle )=\frac{1}{2}|\uparrow\rangle_z\otimes(|ES\rangle+|GS\rangle ).
\label{eq: analyzer}
\end{equation}
The norm then becomes $\frac{1}{4} \,_z\langle\uparrow|\uparrow\rangle_z(\langle ES|ES\rangle+\langle ES|GS\rangle+\langle GS|ES\rangle+\langle GS|GS\rangle )=\frac{1}{2}$, because the excited and ground states are orthogonal ($\langle ES|GS\rangle=\langle GS|ES\rangle=0$). In addition, half of the time, the atom exiting the + exit of the vertical analyzer will be in the ground state and half of the time in the excited state. The analysis for the -- exit yields identical final probabilities.

Next, we would like to erase the which-way information and restore the initial spin state the atom had after it emerges from the analyzer loop-exciter. In other words, we want to untag the tagged atoms. This requires two stages to work. First, we must have all atoms that emerge from the analyzer loop go through a superpositioner (graphically denoted with an S label), which is described next. This device is called a Hadamard gate in quantum information and is called a $\pi/2$ pulse in nuclear magnetic resonance. We call it a superpositioner, because it corresponds to half of the exciter  operation, which creates a superposition of ground and excited states. In other words, it tranforms the ground state to the superposition $|GS\rangle\rightarrow \frac{1}{\sqrt{2}}(|GS\rangle+|ES\rangle)$ and it transforms the excited state to the superposition $|ES\rangle\rightarrow \frac{1}{\sqrt{2}}(|GS\rangle-|ES\rangle)$. Because these two states are orthogonal to each other, we can still differentiate them, so the superpositioner does not erase the which-way information. But the superpositioner does change the quantum state. This is a different operation from a simple change of basis.

\begin{figure}[htb]
\centering
\includegraphics[width=4.5in]{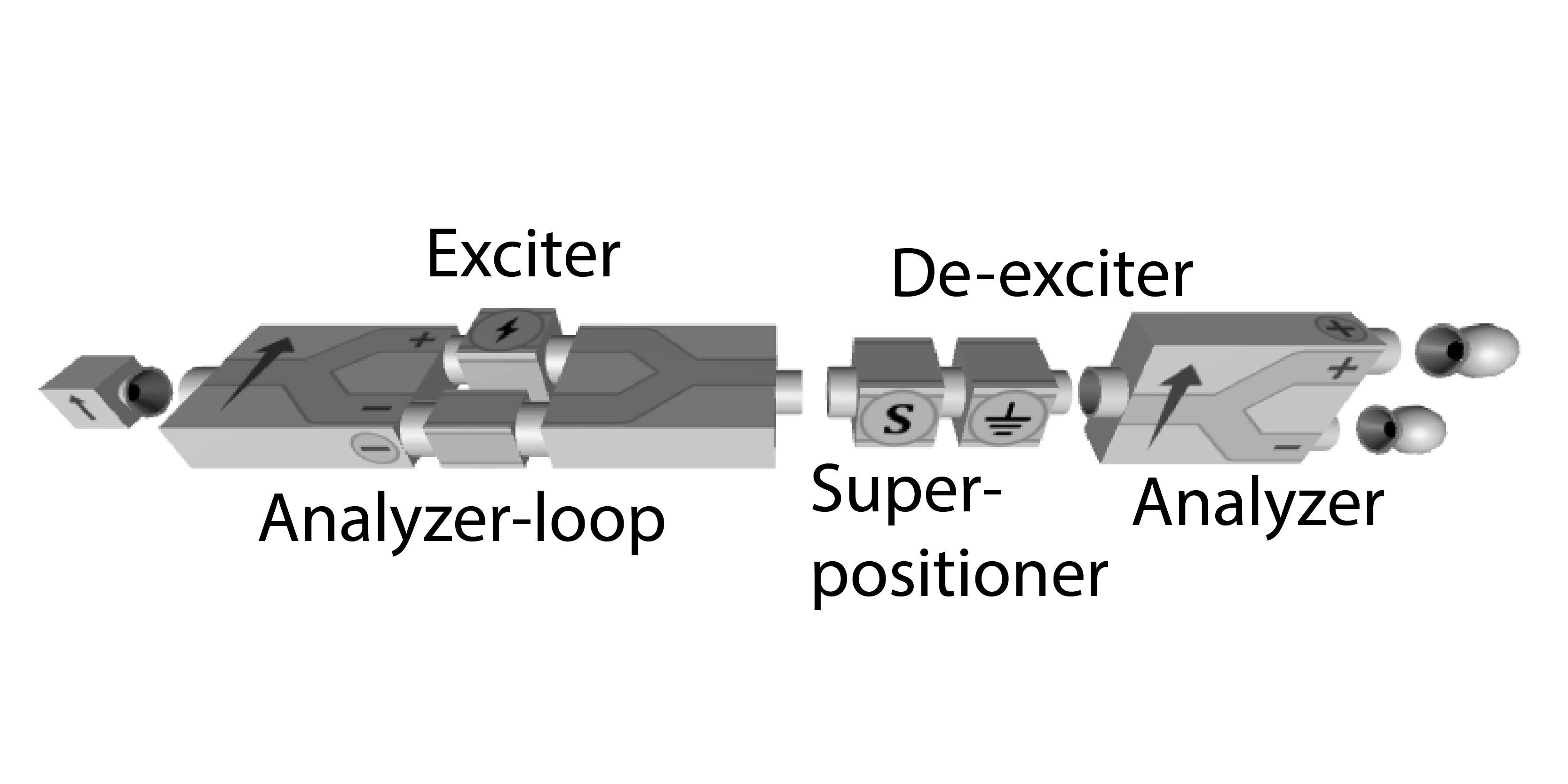}\\
\includegraphics[width=4.5in]{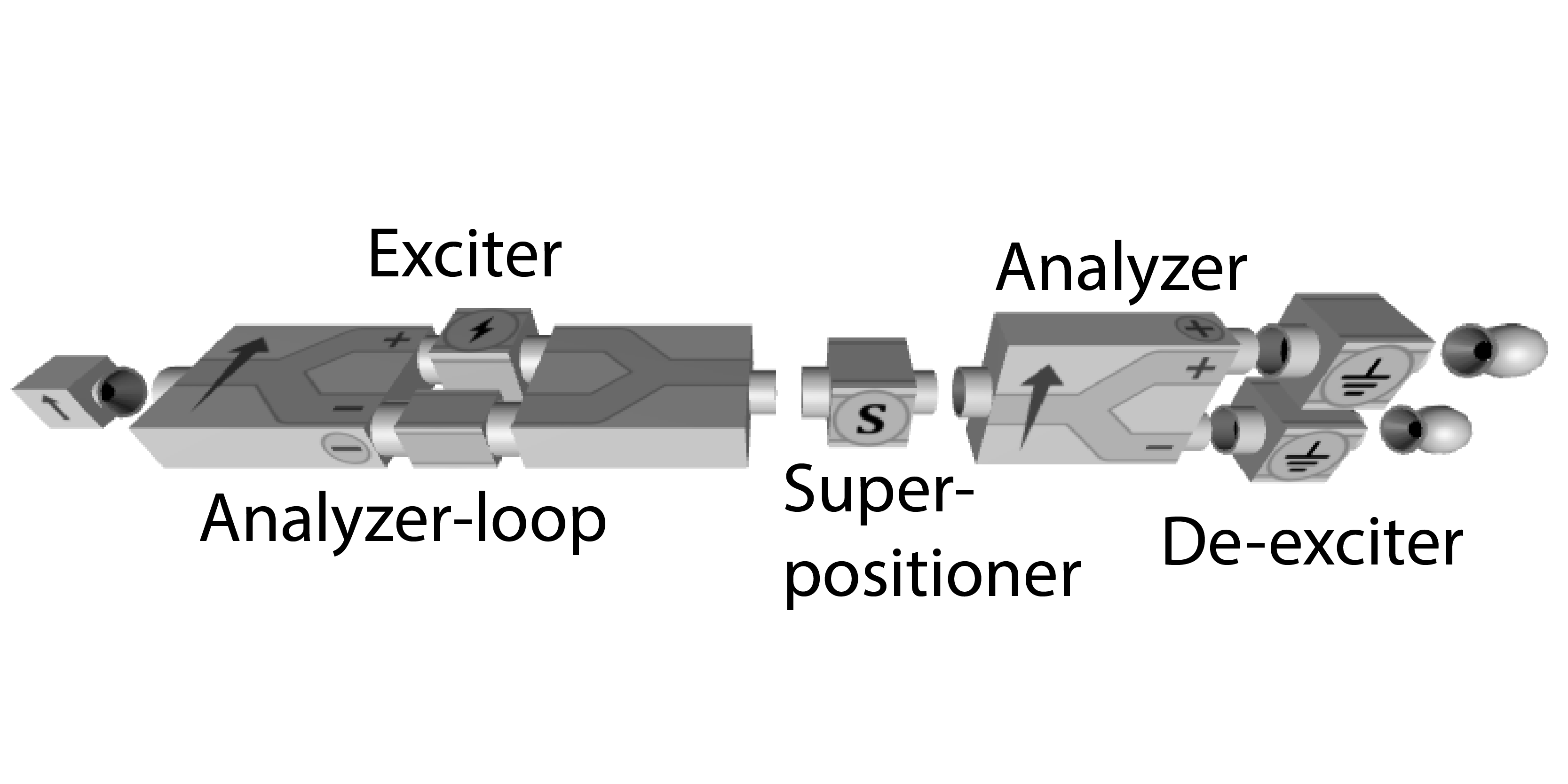}
\caption{ Full eraser experiment with the eraser elements (superpositioner and de-exciter) either both positioned before the final analyzer (top) or one before and one after (bottom). In the second case, the de-exciter can be placed as far from the analyzer as desired. }
\label{fig: full_eraser}
\end{figure}

The which-way information is erased by measuring only the atoms in the ground state by employing a de-exciter (denoted by the electrical ``ground'' symbol). The de-exciter will force the excited state to transition to the ground state and emit a photon, but it does nothing to the ground state. If a photon is detected, the de-exciter blocks the atom and does not allow it to exit. Hence, the de-exciter acts like a pass-through filter, which only allows ground-state atoms to pass through; hence, an equivalent name would be ``ground-state filter.'' We can perform this measurement any time before the atom enters the analyzer or after the atom has emerged from an exit of the vertical analyzer (see Fig.~\ref{fig: full_eraser}). This allows us to make a delayed choice for whether we erase the quantum information or not. And the choice can be made {\it after all other measurements have been completed!}

The analysis is completed as follows: Begin with the state emerging from the analyzer loop-exciter, given by $\frac{1}{\sqrt{2}}(\uparrow\rangle_x\otimes |ES\rangle+|\downarrow\rangle_x\otimes |GS\rangle)$. After the superpositioner, this state becomes 
\begin{equation}
\frac{1}{2}[(|\uparrow\rangle_x+|\downarrow\rangle_x)\otimes|GS\rangle+(-|\uparrow\rangle_x+|\downarrow\rangle_x)\otimes|ES\rangle].
\end{equation}
Changing to the $z$-basis for the spin, we find, the state after the superpositioner is $\frac{1}{\sqrt{2}}(|\uparrow\rangle_z\otimes|GS\rangle-|\downarrow\rangle_z\otimes |ES\rangle)$. Hence, we have shifted the entanglement to now be the superposition of an up spin along the $z$-axis correlated with the ground state and the down spin along the $z$-axis correlated with the excited state. Now, if we decide to record the measurements only for atoms that emerge from the de-exciter (in the ground state), we have erased the which-way information, and we find the atom emerges from the analyzer with the same state it entered the analyzer loop, namely the positive projection of spin along the $z$-axis. 

Note that we lose half of the atoms and half of the measurements when we do this. This behavior is typical of quantum eraser measurements. We must remove the atoms that have the wrong quantum behavior and hence we lose signal when we restore the original quantum coherence that we lost by determining the which-way information. While, in principle, one might be able to devise a clever way to overcome this issue by using interaction-free measurements, it appears to be an issue with all quantum eraser measurements. The full quantum state is not restored by the eraser, because we must remove the ``bad'' measurements from the experiment.  Note, on the other hand, if we do not measure the internal state of the final atom, then we find half of the atoms emerge from the + exit and half from the -- exit. This is what happens when the atoms are watched, or whenever we have which-way information.

Wheeler originally suggested that perhaps the delayed choice measurement implies that the quantum particles infer their behavior by moving backwards in time. But we see this is not necessary at all when one performs a careful analysis. Indeed, this effect arises solely from the correlations and entanglement between the different quantum states of the quantum particle (ground or excited state and spin). Similarly, in a two-slit experiment it arises from which slit the photon went through and its polarization. Hence, all of the information is in the linear combinations of direct products of the wavefunction and that is all one needs to understand and analyze these experiments.

There are a number of variants one can include for further discussion or as problems for the students. These include the following possibilities: (1) change the orientation of the analyzer loop from a horizontal direction to a different angle with respect to the vertical such as 45 degrees; (2) place the de-exciter in front of the final vertical analyzer, so that all of the atoms that emerge from the final analyzer are ground state atoms in the + state along the $z$-axis;  (3) allow the students to complete the delayed choice analysis instead of doing it for them and (4) have the sutdents discuss whether the superpositioner could be placed after the vertical analyzer but before the de-exciter.

In addition to providing a neat exercise in working with direct product states, the analysis of the delayed choice Stern-Gerlach quantum eraser allows the students to fully understand a complex experiment with a rather elementary analysis, which requires applying just a few quantum rules. When coupled with videos of the quantum eraser for the two-slit experiment, this can be a powerful way to help students understand quantum phenomena early in the curriculum and to build confidence that this material can be understood easily if one simply analyzes the behavior according to the quantum rules.

\section{Possible implementation in a real atomic system}

The main challenge with implementing the delayed choice Stern-Gerlach quantum eraser in a real system is that the transition between the internal states of the atom must not change the total electronic angular momentum of the system, which determines the projection of the angular momentum onto the axis of the Stern-Gerlach device. Electronic transitions between different atomic energy levels are likely to affect such states as the total angular momentum usually changes for these transitions.  Furthermore, such excited states are very short-lived (few ns to $\mu$s), and would not survive long enough for an experiment to be completed.

Instead, we propose another type of system, which has a good potential to work, but may be difficult to implement in practice. The system is the $^{171}$Yb atom, a species known to enable an ultra-accurate optical frequency atomic clock.\cite{lemke}~~The Ytterbium atom has two J = 0 atomic clock states, the $^1$S$_0$ and the $^3$P$_0$ states, each of which has angular momentum zero. The $^{171}$Yb isotope also has a nuclear spin one-half, and can be prepared and detected in either its positive or negative projection states.  Although the $^1$S$_0\rightarrow ^3$P$_0$ clock transition near a laser wavelength of 578 nm is strictly forbidden, the presence of the nuclear spin breaks the symmetry and permits laser excitation to the excited state, so that any superposition of ground and excited states could be prepared in the atomic clock experiment.   Since the coupling of these $J=0$ electronic states to the nuclear spin is extremely weak, the excited state lifetime is quite long, and the nuclear spin constants are nearly the same in the ground and excited states.  Thus, the electronic and nuclear spin degrees of freedom can be taken as essentially independent.

While one might think that the $^{171}$Yb  atom provides a nearly ideal system to realize our various Stern-Gerlach schemes, there is one problem.  The nuclear magnetic moment,\cite{stone} 0.49367$\mu_N$ for $^{171}$Yb, is much smaller than the electron magnetic moment used for a typical Stern-Gerlach separation of spin states.  Electronic magnetic moments are on the order of one Bohr magneton ($\mu_B/\hbar$ = 14.0 GHz/T), whereas the nuclear magneton ($\mu_N/\hbar$ = 7.62 MHz/T) is nearly 2000 times smaller.  The original experiment of Stern and Gerlach used a beam of silver atoms, which have a single unpaired electron.  They were able to separate the two electronic spin projections by several tenths of a mm using a quite strong field gradient of a few T/cm.  Thus, achieving practical separations with a small nuclear magnetic moment requires impractically large magnetic field gradients.  This certainly creates a challenge with implementing such an experimental system in practice, but it does show that in principle, such a system can be used in these thought experiments.

It may be possible to use the optical Stern-Gerlach (OSG) effect to achieve large enough separations in order to implement our scheme.  The separation of nuclear spin components using the OSG method has already been demonstrated\cite{osg1} with $^{171}$Yb and $^{173}$Yb and the similar atomic clock species\cite{osg2} $^{87}$Sr.  The latter species has nuclear spin of 9/2, which could be separated into 10 separate spin projection states using the OSG effect with ultracold atoms.  The optical separation is based on using the strong light intensity gradient in a focused laser beam to separate the different spin components, which couple differently to the laser field and experience differential optical forces.  Whether a practical OSG experiment could be designed for our scheme would need to be carefully considered, since the ground and excited electronic states do not in general experience the same optical forces, although it is often possible to find “magic wavelengths” where they are the same.

\section{Application to Other Experiments}

One of the most common examples of a delayed choice quantum eraser is to perform the two-slit experiment with crossed polarizers over the slits and a polarizer that is employed at the screen before measuring the pattern of light. If the polarizers at the slits are horizontal and vertical, respectively, then a horizontal polarizer at the screen will see a single slit pattern, as will a vertical polarizer. But if it is rotated to 45 degrees, then the interference pattern emerges. Numerous YouTube videos of this experiment exist, and it can be implemented rather easily at home using just a laser pointer and polarizers from 3D movie glasses. 

Because this paper is focused on the Stern-Gerlach experiment, we do not go through the full analysis of the conventional two-slit experiment here, but it should be clear that a quite similar analysis can be done of this experiment, and it reinforces the concepts covered for the Stern-Gerlach experiment. Depending on when one wants to discuss polarization in the quantum mechanics class, this might come later in the curriculum than the Stern-Gerlach experiment.

In addition, the same techniques employed here for the delayed choice Stern-Gerlach experiment can also be employed to examine  other interesting experiments, as Styer does in his text. These include a modified version of the Einstein-Podolsky-Rosen experiment and of the Bell experiments. We feel including all of these additional topics greatly enhances the undergraduate quantum curriculum and would not take too much time away from more standard topics. But we feel the benefits that the student gains from having contact with modern quantum experiments and from understanding concepts such as superposition and measurement in a more concrete fashion far outweigh the cost in time to other subjects which might need to be dropped from the course.

\section{Conclusions}

As more and more quantum classes embrace the Stern-Gerlach-first curriculum, it becomes possible to employ this experiment to cover a range of interesting modern quantum experiments that showcase the bizarre nature of quantum mechanics while strengthening the students' abilities in understanding concepts  such as superposition, direct products,  and measurement. Tackling these concepts early on will help ground the students in the fundamentals of quantum mechanics and better prepare them for the rest of the quantum curriculum they will cover in their course. Given the fact that they already have all of the prerequisite knowledge needed from current textbook coverage of the Stern-Gerlach experiment, we owe it to our students to provide them with this entry into more sophisticated material. We hope other quantum mechanics instructors will agree.

\begin{acknowledgments}

This work was supported by the National Science Foundation under grant number PHY-1620555. In addition, J.K.F. was also supported by the McDevitt bequest at Georgetown University.

\end{acknowledgments}

\end{document}